\documentclass[a4paper]{JHEP3}

\usepackage[latin1]{inputenc}
\usepackage{bbm}
\usepackage{bm}
\usepackage{graphicx}
\usepackage{epsfig}
\usepackage{subfigure}
\usepackage{latexsym}
\usepackage{amsmath}
\usepackage{amsfonts}
\usepackage{amssymb}

\newcommand{\Tr}{\textrm{Tr}}

\newcommand{\ev}[1]{\langle #1 \rangle}

\newcommand{\nr}[1]{(\ref{#1})}

\newcommand{\Dslash}{\ensuremath \raisebox{0.025cm}{\slash}\hspace{-0.28cm} D}
\newcommand{\be}{\begin{eqnarray}} 
\newcommand{\ee}{\end{eqnarray}}

\newcommand{\bmp}{\noindent\begin{minipage}{16cm}}
\newcommand{\emp}{\end{minipage}\vskip 7mm} 
\def\lsim{\mathrel{\raise.3ex\hbox{$<$\kern-.75em\lower1ex\hbox{$\sim$}}}}
\def\gsim{\mathrel{\raise.3ex\hbox{$>$\kern-.75em\lower1ex\hbox{$\sim$}}}}

\newcommand{\intron}[1]{}

\title{Spectrum of SU(2) lattice gauge theory with two adjoint Dirac flavours}

\author{Ari J. Hietanen\\ Department of Physics, Florida International University}
\author{Jarno Rantaharju, Kari Rummukainen\\ Department of Physics, University of Oulu}
\author{Kimmo Tuominen\\ Department of Physics, University of Jyv\"askyl\"a and Helsinki Institute of Physics, University of Helsinki, Finland}
\abstract {An SU(2) gauge theory with two fermions transforming under
  the adjoint representation of the gauge group may appear conformal
  or almost conformal in the infrared. We use lattice simulations to
  study the spectrum of this theory and present results on the masses
  of several gauge singlet states as a function of the physical quark
  mass determined through the axial Ward identity and find indications
  of a change from chiral symmetry breaking to a phase consistent with
  conformal behaviour at $\beta_L\sim 2$. However, the measurement of
  the spectrum is not alone sufficient to decisively confirm the
  existence of conformal fixed point in this theory as we show by
  comparing to similar measurements with fundamental fermions. Based
  on the results we sketch a possible phase diagram of this lattice
  theory and discuss the applicability and importance of these results
  for the future measurement of the evolution of the coupling
  constant.  } 

\begin{document}


\section{Introduction and summary}

Different phases of non-Abelian gauge field theories are already manifest in the Standard model of elementary particle interactions, and charting the phase structure of these theories as, for example, numbers of colours and quark flavours are varied is important for model building beyond the Standard model, see e.g. \cite{Sannino:2008ha}. In this regard, an especially interesting class of theories are quantum field theories with nontrivial infrared fixed points of the $\beta$-function. This means that while the coupling runs when probed at very short distances, it becomes a constant over some energy range in the infrared and the theory appears conformal. One of the phenomenological connections to these theories was provided in \cite{Georgi:2007ek} where the possibility of a fully conformal sector coupled only weakly to the Standard Model through effective operators at low energies was considered. Conformal symmetry determines the mass dimensions of these operators which have some striking implications for the related low energy observables \cite{Georgi:2007si}. Various further phenomenological and theoretical aspects of these unparticles have been investigated, see e.g. \cite{Cheung:2007zza}.

Another phenomenological motivation to study theories which either feature an infrared fixed point or are, in theory space, close to one which does, originates from technicolor (TC) and the associated extended technicolor (ETC) models.  These models were devised a long time ago to explain the mass patterns of the Standard Model gauge bosons and fundamental fermions without the need to introduce a fundamental scalar particle \cite{TC,Eichten:1979ah,Hill:2002ap}. The Higgs sector in these theories consists of a new fermion species (techniquarks), charged under a new gauge interaction (technicolor). Early TC models, based on a technicolor sector straightforwardly extrapolated from a QCD-like strongly interacting theory, had several problems. These include flavour changing neutral currents due to the extended technicolor interactions and unwanted additional light pseudo-Goldstone bosons due to the breaking of the chiral symmetry of the techniquarks. It has been known for some time that these problems are solved in so called walking technicolor theories \cite{Holdom:1981rm,Yamawaki:1985zg,Appelquist:an}. These theories are nearly conformal, i.e. the evolution of the coupling constant is, in a wide range of energy, governed by an attractive quasi-stable infrared fixed point at strong coupling. 

\begin{figure}
\begin{center}
\includegraphics[width=0.6\textwidth]{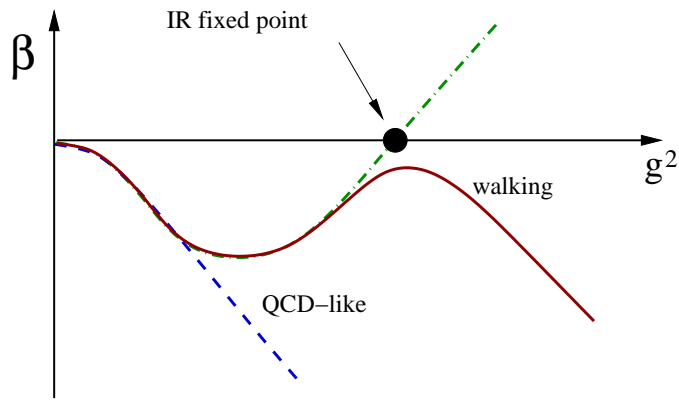}
\caption[a]{The schematic $\beta$-function of a theory with an infrared   fixed point (dash-dotted line, top), walking coupling (solid line,   middle) and QCD-like running coupling (dashed line, below).}
\end{center}
\label{walk_beta}
\end{figure}

Several possible generic forms of a $\beta$-function are sketched in Fig.~\ref{walk_beta}. 
The form of the full nonperturbative $\beta$-function of a non-supersymmetric SU($N$) gauge field theory remains unknown to 
date\footnote{However, for a recent conjecture about its possible form, see \cite{Ryttov:2007cx}}. Nevertheless, we can discuss the features shown in Fig.~\ref{walk_beta} by considering the perturbative $\beta$-function for SU($N$) gauge field theory with fermions, defined up to two loop order as
\begin{equation}
\beta(g)=-\beta_0\frac{g^3}{16\pi^2}-\beta_1\frac{g^5}{(16\pi^2)^2},
\end{equation}
where the coefficients are
\begin{align}
\beta_0&=\frac{11}{3}C_2(G)-\frac{4}{3}T(R)N_f, 
\nonumber \\ \beta_1&=\frac{34}{3}C_2(G)^2-\frac{20}{3}C_2(G)T(R)N_f-4C_2(R)T(R)N_f,
\nonumber
\end{align}
and $C_2(R)$ is the second Casimir invariant for SU($N$) representation $R$ and $\delta^{ab}T(R)={\textrm{tr}}(T^a_RT^b_R)$. As is well known for QCD, for modest amount of matter in the fundamental representation there is not enough screening to compensate for the antiscreening contribution of the gluons and $\beta(g)<0$, which guarantees asymptotic freedom. Perturbative $\beta$-function becomes zero only in the ultraviolet fixed point at $g=0$. However, imagine adding enough matter in the fundamental representation to make 
$\beta_0$ small and positive, while $\beta_1$  becomes negative. Then there appears a nontrivial zero, $\beta(g^\ast)=0$ at $g^{\ast 2}=-16\pi^2\beta_0/\beta_1$, and the $\beta$-function is of the form shown by the dashed-dotted curve in Fig.~\ref{walk_beta}. This zero corresponds to an attractive infrared fixed point. If approached from the asymptotically free side the theory becomes conformal, very unlike QCD, when long-wavelength probes are considered. If $g^\ast$ is small, perturbative analysis can be applied \cite{Banks:1981nn}. However, it is more likely that such infrared fixed points do not appear at perturbative values of the coupling and one has to account for full nonperturbative dynamics of non-abelian gauge theories. This can lead to the walking dynamics.

Namely, one has to take into account the formation of quark-antiquark condensate leading to the spontaneous breaking of chiral symmetry, which decouples the quarks from infrared dynamics and keeps the $\beta$-function negative. 
Therefore the $\beta$-function never actually reaches the would-be infrared fixed point, but the very close proximity to it allows the coupling to ``walk'' slowly over a wide range of scales.  Eventually the $\beta$-function is repelled from the proximity of the fixed point and behaves as in a generic asymptotically free theory, see Fig.~\ref{walk_beta}. Another way to destabilize the infrared fixed point is to add small explicit chiral symmetry breaking, e.g. a mass term for the fermions.  

Therefore, there is need for concrete examples of theories with infrared stable fixed points but not many concrete examples are currently known. For example ${\mathcal{N}}=4$ SYM provides a concrete realization of a theory where the $\beta$-function vanishes identically as was first shown on the two loop level in \cite{Jones:1977zr}, and later to all orders in \cite{Brink:1982wv}. This theory, however, is conformal at all scales and not just in the infrared. Supersymmetry provides enough structure to allow for nonperturbative analysis and construction of theories which feature an infrared stable fixed point at which the $\beta$-function becomes zero and the running coupling freezes to a constant value. Beyond supersymmetric cases \cite{Intriligator:1995au}, little is known in full quantitative detail, and the reason is that this requires nonperturbative knowledge which to date can only be obtained on the lattice.

On the other hand, analytic semi-quantitative studies of conformal windows have been able to single out some viable candidates which should be studied in more detail using lattice methods. In particular, for non-supersymmetric Yang-Mills theories with higher fermion representations it has been suggested \cite{Sannino:2004qp} that an ideal candidate for minimal walking technicolor theory would be the one with just two (techni)quark flavours in two-index symmetric representation of SU(2) or SU(3). Initial studies of these theories have been performed on the lattice already \cite{Catterall:2007yx,Catterall:2008qk,Shamir:2008pb}. For a related study of a QCD-like theory with fundamental representation fermions see \cite{Appelquist:2007hu}.

In this paper we consider the case of SU(2) gauge fields with two fermions in the two-index symmetric representation, which, for SU(2), is equivalent to the adjoint representation.  We measure various correlators as functions of the fermion mass and lattice coupling in order to obtain information on the spectrum of (techni)color singlet states of this theory. 

Naturally, the 
most direct 
way to answer the question concerning the conformal behaviour of the coupling constant in this theory would be to directly measure the evolution of the coupling constant.  Although this can be implemented on the lattice straightforwardly, using e.g. the Schr\"odinger functional method, care is needed in the interpretation of the results. However, even without direct knowledge of the behaviour of the coupling constant one can learn about the possible conformal behaviour through other observables. In the conformal theory, no massive bound states can exist, since the existence of a mass scale immediately implies breaking of the conformal invariance. Hence, the spectrum of color singlet states should provide first hints of the possible underlying conformal invariance. Furthermore, 
regarding
the possible conformal or walking behaviour it is important to understand the onset of chiral symmetry breaking in these theories. To complement the initial studies \cite{Catterall:2007yx}, we present our high-precision results  concerning the spectrum of the color singlet states in the SU(2) gauge theory with two adjoint fermion flavours and discuss the implications this will have on the possible conformal (or walking) behaviour. We will show that while our results seem consistent with near conformal behaviour, the definite answer to this question can only be provided in a further study. 

The paper is structured so that in sec.~\ref{model} we recall the basics of the model as well as of the lattice formulation we use, in sec.~\ref{results} we present our results and interpretation of the data and in sec.~\ref{outlook} we conclude and outline the directions of our future work. 

\section{The model and lattice formulation}
\label{model}

The minimal model to be studied here consists of the gauge dynamics of two Dirac fermions in the adjoint representation of the SU(2) gauge theory. This model has been proposed to be (quasi)conformal with only two flavours of massless technifermions.  

The continuum theory is defined by:
\begin{eqnarray}
	{\mathcal{L}}=-\frac{1}{2}F_{\mu\nu}F^{\mu\nu}+
        \overline{\textrm{u}}~\Dslash \textrm{u} +
        \overline{\textrm{d}}~\Dslash \textrm{d}
\end{eqnarray}
where $F_{\mu\nu}$ is the usual SU(2) field strength, and the
gauge covariant derivative is
\begin{eqnarray}
	\left[D_\mu \textrm{u}\right]^a =
       \left(\partial_\mu\delta^{ac}-ig_{\textrm{TC}}A^b_\mu\epsilon^{abc}\right)
       \textrm{u}^c,
\end{eqnarray}
where $a=1,2,3$. 

Due to the pseudoreality of the SU(2) representations, the global symmetry group of the above theory is SU(4), which breaks to ${\textrm{SO}}(4)$ due to a formation of the fermion condensate $\langle\overline{\textrm{u}}\textrm{u} + \overline{\textrm{d}}\textrm{d}\rangle$. This pattern leads to the appearance of nine Goldstone bosons. If the above theory is taken to drive the electroweak symmetry breaking, i.e. coupled to the electroweak SU$_{\textrm{L}}$(2)$\times$U$_{\textrm{Y}}$(1) gauge fields, three of these Goldstones will become the longitudinal components of the weak gauge bosons, and the physical low energy spectrum is expected to contain six Goldstone bosons and these are furthermore expected to receive masses of the order of the electroweak scale through, e.g. ETC interactions. Here we are interested of the strong coupling properties of the SU(2) gauge theory and do not consider the coupling to the electroweak.

We will study the masses of several (techni)color neutral states. In the adjoint theory the spectrum is much richer than if fundamental representation fermions are considered. Because the technigluons, quarks and antiquarks all transform in the same representation, we can obtain gauge singlet states by combining two or more of any of the fields above. The simplest to measure are meson-like states consisting of two quarks 
or a quark and an antiquark. The three quark color neutral states can be constructed in analogy to baryons of ordinary QCD, e.g. ``proton'' (spin-1/2) and ``delta'' (spin-3/2). Finally there are possible exotic states like the color neutral combination of a techniquark and technigluon.   

On the lattice the action is
\begin{eqnarray}
	S_{\textrm{lat}}=S_G+S_F,
\end{eqnarray}
with the standard Wilson plaquette action
\begin{equation}
  S_G = \beta_L \sum_{x;\mu<\nu} 
  \left [1 - \frac12 \Tr P_{x;\mu\nu}\right],
\end{equation}
where $P_{\mu\nu}$ is the standard $1\times 1$ plaquette,
written in terms of the SU(2) fundamental representation link matrices $U_\mu(x)$.  The Wilson fermion action, $S_F$, for two (degenerate) Dirac fermions in the adjoint representation of the gauge group is
\begin{equation}
  S_{\textrm{F}} = \sum_{f=\textrm{u},\textrm{d}} 
  \sum_{x,y} \bar\psi_{f,x} M_{xy} \psi_{f,y} ,
  \label{sf}
\end{equation}
where 
\begin{equation}
  M_{xy} = \delta_{xy} - 
  \kappa \sum_\mu \left[ (1+\gamma_\mu) V_{x,\mu} +
         (1-\gamma_\mu)V^T_{x-\mu,\mu}\right].
\end{equation}
The only modification to the usual Wilson action for fundamental fermions is the replacement of the link variables $U_\mu(x)$ with the variables
\begin{equation}
	V_\mu^{ab}(x)=2{\textrm{tr}}(S^aU_\mu(x)S^bU_\mu^\dagger(x)),
\end{equation}
where $S^a$ $a=1,2,3$ are the generators of the fundamental representation, normalised as $\Tr S^aS^b=\frac12 \delta^{ab}$.
The elements of $V$-matrices are real and $V^{-1} = V^T$.  
We note that because the link matrices are real, it would be possible to perform lattice simulations with two Dirac flavours of staggered fermions without encountering the square root problem (for an example, see \cite{Karsch:1998qj}).

The lattice action is parametrised with two dimensionless parameters,
$\beta_L = 4/g_{\textrm{bare}}^2$ and 
$\kappa = 1/[8+2(am_{q,\textrm{bare}})]$.
The parameter  $\kappa$ is related to the quark mass.   Because the Wilson fermion action breaks the chiral symmetry of the (massless) continuum theory, the bare quark mass receives large additive corrections, and we cannot avoid the inclusion of the mass term in the lattice action.  The physical (PCAC) quark mass is defined through the flavour non-diagonal axial Ward identity
\begin{equation}
  m_{\textrm Q}=\lim_{t \rightarrow \infty}
  \frac{1}{2}\frac{\partial_t V_{\textrm{AP}}}{V_{\textrm{PP}}},
  \label{pcac}
\end{equation}
where the pseudoscalar-pseudoscalar and axial-pseudoscalar currents
are as follows:
\begin{align}
  V_{\textrm{PP}}(t) &= \sum_{\bm{x},\bm{y}} \ev{ 
    \bar{\textrm{d}}(\bm{x},t)\gamma_5 \textrm{u}(\bm{x},t)
    \bar{\textrm{d}}(\bm{y},0)\gamma_5 \textrm{d}(0,0)} \\
  V_{\textrm{AP}}(t) &= \sum_{\bm{x},\bm{y}}
    \ev{\bar{\textrm{d}}(\bm{x},t)\gamma_0\gamma_5
      \textrm{d}(\bm{x},t)
       \bar{\textrm{d}}(\bm{y},0)\gamma_5\textrm{d}(0,0)}.
\end{align}
Here and below we use wall sources with Coulomb gauge fixing in all
our hadronic correlation functions.
For each $\beta_L $ the value of the hopping parameter $\kappa$ where $m_Q$ vanishes defines the critical hopping parameter $\kappa_c(\beta_L)$.  In practice we use a few $t$-values around $t=L_t/2$ for the fit in Eq.~\nr{pcac}.

The Wilson quark action \nr{sf} has $O(a)$ discretisation errors, which could be eliminated by employing the $O(a)$ improved ``clover'' action.  However, to fully implement $O(a)$ improvement would substantially increase the complexity of the simulation program and require careful renormalisation of various quantities.  Thus, the lowest-order accuracy is justified in exploratory investigation as is the case here.
The simulations are performed with the hybrid Monte Carlo algorithm, using even-odd preconditioned quark matrix and leapfrog integrator.  The quark matrix is inverted using the standard conjugate gradient method.

The masses of colour singlet ``hadrons'' are estimated by fits to the time sliced averaged correlation functions with Coulomb gauge fixed wall sources.  We concentrate only on flavour non-diagonal (isospin non-singlet) operators.  For example, the correlation function for quark-antiquark states, ``mesons,'' is given by
\begin{equation}
   G_X(t)=\sum_{\bm{x},\bm{y}}
   \langle\bar{\textrm{u}}(\bm{x},t)\Gamma_X \textrm{d}(\bm{x},t)
   \bar{\textrm{d}}(\bm{y},0)\Gamma_X\textrm{u}(0,0)\rangle,
   \label{meson}
\end{equation}
where $\Gamma_X = \gamma_5$ for the pseudoscalar and $\Gamma_X =\gamma_i$, $i=1,2,3$ for the vector meson.  We also measured scalar ($\Gamma_X = 1$) and axial vector ($\Gamma_X = \gamma_5\gamma_i$) correlators, but for these the statistical errors remain too large for reliable results.  Except for the isosinglet channel, states consisting of two quarks are degenerate with the quark-antiquark states in Eq.~\nr{meson}.  

The correlation functions for three quark states, spin-1/2 ``proton'' and spin-3/2 ``$\Delta$'', are constructed in analogous fashion to standard SU(3) QCD.  We also made an attempt to measure the correlation function of the colour singlet quark-gluon state; however, we did not succeed in constructing a source operator without  
excessive statistical noise for measuring the mass.

\section{Results}
\label{results}

The simulations were carried out with five different values of $\beta_L=1.3$, 1.7, 1.9, 2.2 and 2.5.  For each value of $\beta_L$ we used 5 to 11 different values of $\kappa$, corresponding to different quark masses, with  volumes $24^4$ and $32^4$.  In addition, we made
exploratory runs with volumes $8^4$ and $10^4$, but we did not use these volumes for spectroscopy.  For each case the number of hybrid Monte Carlo trajectories was 100-700. The evolution timestep $\Delta \tau$ was $0.02$ for larger values of the quark mass and was decreased to $0.003$ closer to the zero mass limit. The number of integration steps $N_s$ was chosen so that the trajectory length $N_s \times \delta\tau \sim \mathcal{O}(1)$.  For comparison we also performed simulations with SU(2) gauge group and two flavours of fundamental representation quarks, using $\beta_L = 1.7$ and 2.5.  Some of the early results of this study have appeared in \cite{Hietanen:2008vc}.

As briefly discussed in the introduction, we are investigating how the phenomena of chiral symmetry in the continuum emerge from the lattice which has no such symmetry to begin with. The standard way is to study the critical line $\kappa_{c}(\beta_L)$, defined as the line where the quark mass (as defined in Eq.~\nr{pcac}) vanishes.  If there is chiral symmetry breaking, along this line the pseudoscalar ``pions''  become massless Goldstone bosons, whereas other hadrons remain massive.  If $m_Q > 0$ ($\kappa < \kappa_c(\beta)$), the mass of the pseudoscalars is $\propto \sqrt{m_Q}$.  Locating this critical line allows us to sketch a possible phase diagram of the lattice theory.

\begin{figure}
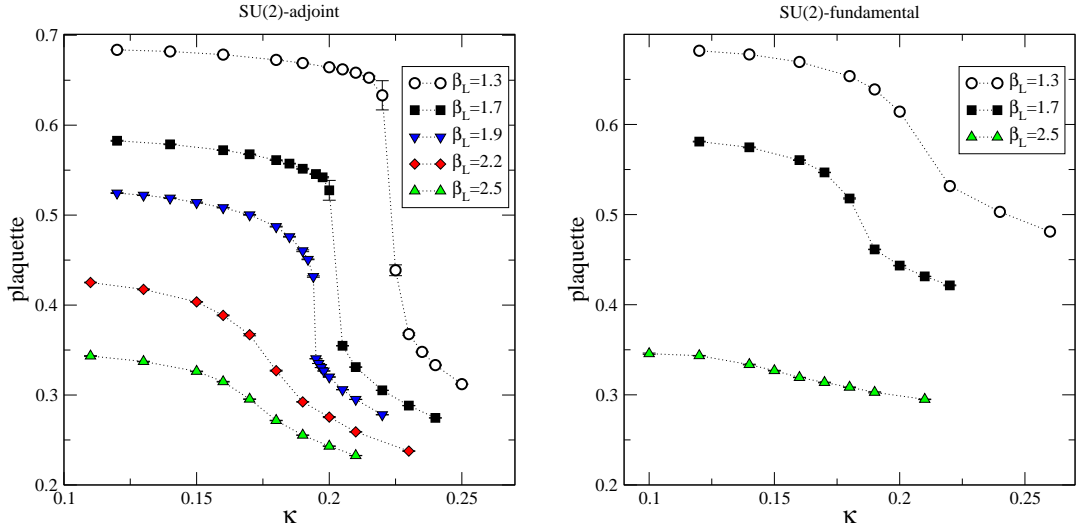

\includegraphics[width=0.45\textwidth]{plaq_su2a_small.eps} ~~~
\includegraphics[width=0.45\textwidth]{plaq_su2_small.eps}
\caption[a]{Left panel: The plaquette expectation value at fixed values of the lattice coupling $\beta_L=4/g_{\textrm{bare}}^2$ as functions of the hopping parameter $\kappa$ for the theory with adjoint representation fermions.  The measurements are done using $10^4$ lattices. Right panel: Same as left, but for the fundamental representation fermions.}
\label{plaq_su2}
\end{figure}

However, in practice the simulations become progressively more difficult the smaller the quark mass is.  This is due to the appearance of very small eigenvalues of the fermion matrix, which finally make the inversion of the matrix impossible.  This becomes more severe in large volumes.  In order to better understand the phase diagram, we performed a set of simulations on a relatively small $10^4$ lattice. In the left panel of Fig.~\ref{plaq_su2} we show the results for the measurement of the plaquette expectation value as a function of $\kappa$ at fixed lattice coupling $\beta_L=4/g_{\textrm{bare}}^2$ in SU(2) gauge theory with adjoint fermions. The results show clear
signs of a discontinuity for values of $\beta_L$ smaller than  $\beta_{L,c}\sim 2$, while at larger couplings we obtain a smooth curve.  Even with this volume it turns out to be very difficult to make simulations at very close proximity to the jump.  On the other hand, at larger values of $\beta_L$ it is possible to do simulations at all values of $\kappa$.  We take this to be an indication of a first order phase transition for $\beta_L < \beta_{L,c} ~\sim 2$,
at which point there possibly is a critical point where the transition ends.  Signs of this kind of behaviour were also seen in \cite{Catterall:2008qk}.

This observation can be contrasted with the case of fundamental representation fermions: we have performed the measurements for the same range of $\beta_L$ in the case of fundamental fermions and the results are shown in the right panel of Fig.~\ref{plaq_su2}. In comparison with adjoint fermions, we find considerably smaller discontinuity (or possibly no discontinuity at all) at small values of $\beta_L$.  We note that a first order phase transition at small values of the lattice coupling $\beta_L$ has also been seen in standard SU(3) lattice QCD with Wilson fermions; for a recent review, see \cite{Jansen:2008vs}. These first order transitions are lattice artifacts, not present in the continuum theory.

\begin{figure}
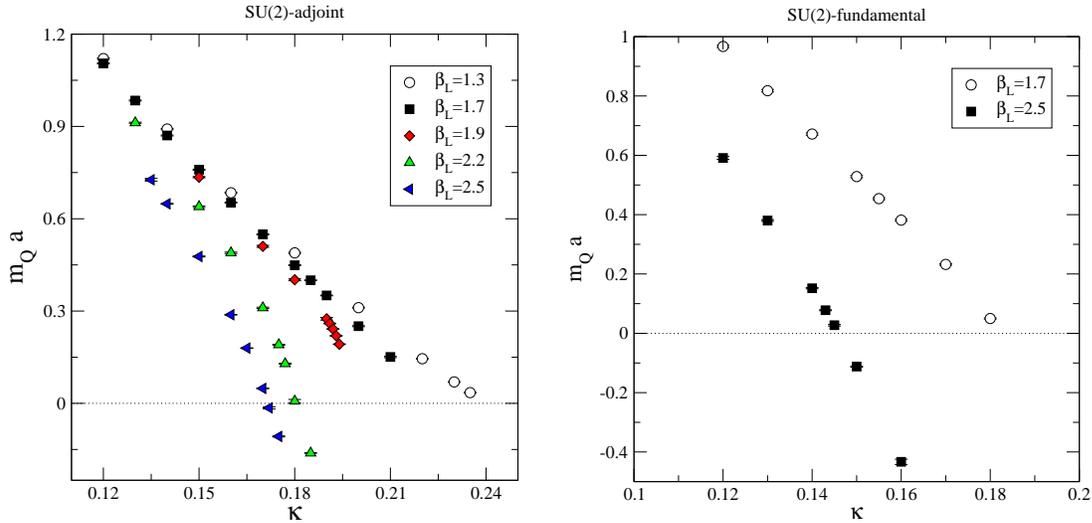

\includegraphics[width=0.45\textwidth]{mq_su2a.eps} ~~~
\includegraphics[width=0.46\textwidth]{mq_su2.eps}
\caption[a]{Left panel: The physical (PCAC) quark mass $m_Q$ as a function of the hopping parameter $\kappa$ for the theory with adjoint representation quarks. Right panel: Same as left but for fundamental representation fermions.}
\label{quarkmass_su2}
\end{figure}

We measure the quark masses and the hadron spectroscopy from larger lattice volumes, $24^4$ or in some cases $32^4$.  In left and right panels of Fig. \ref{quarkmass_su2} we show $m_Q$, measured using Eq.~\nr{pcac}, as functions of the hopping parameter $\kappa$ for different values of $\beta_L$ for the adjoint representation quarks and for the fundamental representation quarks, respectively.  For the adjoint representation, at small $\beta_L \lsim 1.9$ we are not able to reach very small quark masses, because the first order phase transition in left panel of Fig.~\ref{plaq_su2} occurs at $\kappa$-values which correspond to non-zero quark masses\footnote{%
However, note that at $\beta_L = 1.3$ we are able to reach $\kappa = 0.235$, which is substantially beyond the location of the transition at $\kappa\approx 0.22$.  This is presumably due to the large metastability of the first order transition at large volume.}
At this point the measured PCAC quark masses abruptly jump to negative values (measured from smaller volumes; not shown on the plot).  This behaviour has been observed to happen also in standard SU(3) lattice QCD with Wilson quarks \cite{Jansen:2008vs}.  On the other hand, at larger $\beta_L$ we observe no particular problems in simulating with $\kappa$-values corresponding even to negative quark masses.   Naturally, simulations at very small quark masses still require very short update step in HMC trajectories.  We observe qualitatively similar behaviour for fundamental representation quarks.

\begin{figure}
\begin{center}
\includegraphics[width=0.5\textwidth]{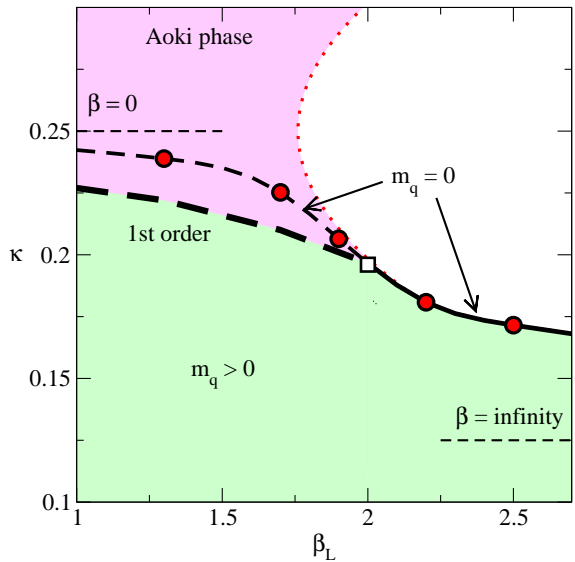}
\caption[a]{The phase diagram on $(\beta_L,\kappa)$ -plane for the adjoint theory.  Solid circles denote the measured (and extrapolated for $\beta_L \le 1.9$) critical hopping parameters $\kappa_c(\beta_L)$, where $m_Q = 0$.  At $\beta_L \lsim 2$ there appears a 1st order phase transition, which ends at a critical point shown by open square.  The phase structure above the critical line is a conjecture, and it can be more complex than shown here.  The 
Aoki phase may exist also for values larger than $\beta_L\sim 2$.}
\end{center}
\label{PD_su2a}
\end{figure}

We are now in position to sketch the phase diagram for the adjoint theory in Fig. \ref{PD_su2a}.  The critical line $\kappa_c(\beta_L)$ is defined as the line where the quark mass vanishes. In the weak coupling limit $\beta_L=0$ it has the value $\kappa=1/4$ and extends towards $\kappa=1/8$ at $\beta_L=\infty$.  In the small-$\beta_L$ -region, where the first order transition prevents us from reaching very small quark masses, we nevertheless extrapolate its location. (This is done only for illustration; we do not need it in subsequent analysis.)  The first order line appears to terminate at a critical point roughly at $\beta_{L,c}\sim 2$,   above which the behaviour appears to be regular as $m_Q = 0$ -limit is passed.  The location of this critical point is conjectural; we cannot numerically exclude a very weak first order transition extending to much higher values of
$\beta_L$, possibly to infinity.  However, the abruptness with which 
the discontinuity vanishes in our simulations may suggest the existence of a genuine critical point.  We remind that the theory
is asymptotically free, thus, the lattice continuum limit is at $\beta_L \rightarrow \infty$, unless there exists a critical point at finite $\beta_L$ leading to non-trivial infrared physics.

We conjecture that the first order transition leads into so-called Aoki phase, which is a strong lattice coupling artifact
for Wilson fermions \cite{Aoki:1986ua,Sharpe:1998xm,DelDebbio:2008wb}. In this phase flavour symmetry and parity are spontaneously broken. Our observations confirm the behaviour established in the earlier studies at smaller volumes \cite{Catterall:2008qk}.  We note that the lattice phase diagram of the theory with fundamental SU(2) representation quarks is expected to have similar features to the adjoint quark one, with the possible exception of the missing first order transition at small $\beta_L$.

\begin{figure}
\begin{center}
\includegraphics[width=0.5\textwidth]{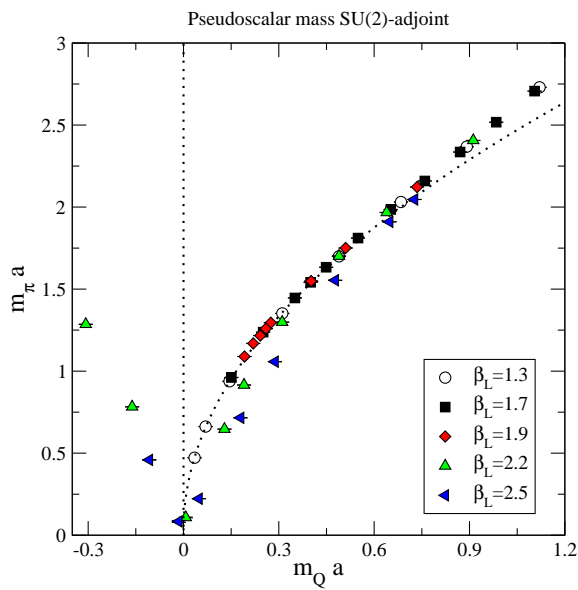}
\caption[a]{Pseudoscalar (``$\pi$'') mass for the adjoint representation quark theory at different values of $\beta_L$. The dotted line is a fit $\propto\sqrt{m_Q}$ to the mass measurements at $\beta_L \le 1.9$ and $m_Qa\le 0.5$.}
\end{center}
\label{pi_su2a}
\end{figure}

\begin{figure}
\begin{center}
\includegraphics[width=0.5\textwidth]{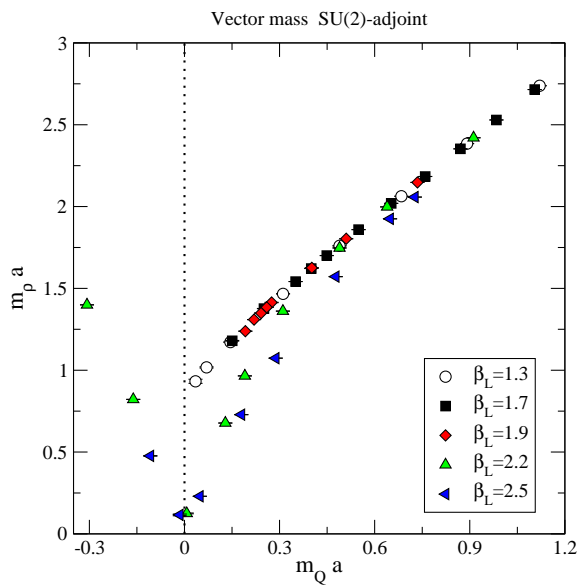}
\caption[a]{Vector meson (``$\rho$'') mass for the adjoint representation quark theory.}
\end{center}
\label{rho_su2a}
\end{figure}

With these results in mind, let us now consider the measurements of the mass spectrum and their implications. 
In Fig. \ref{pi_su2a} we show the pseudoscalar quark-antiquark meson ($"\pi"$) mass as a function of the PCAC quark mass $m_Q$. For strong bare coupling, i.e. small $\beta_L$, we find that the pseudoscalar mass is proportional to $\sqrt{m_Q}$.  This is the expected behaviour in the presence of chiral symmetry breaking, if the pseudoscalar is the Goldstone boson of the broken symmetry.  Consistently with this behaviour the vector meson (``$\rho$'') mass has a finite intercept for these same $\beta_L$ values; the results for the vector mass are shown in Fig.~\ref{rho_su2a}.  However, as $\beta_L$ is increased above $\beta_L\sim 2$, the location of the possible critical point, there is a qualitative change: the pseudoscalar and vector meson masses now appear linearly proportional to $m_Q$, and very nearly degenerate.

\begin{figure}
\begin{center}
\includegraphics[width=0.5\textwidth]{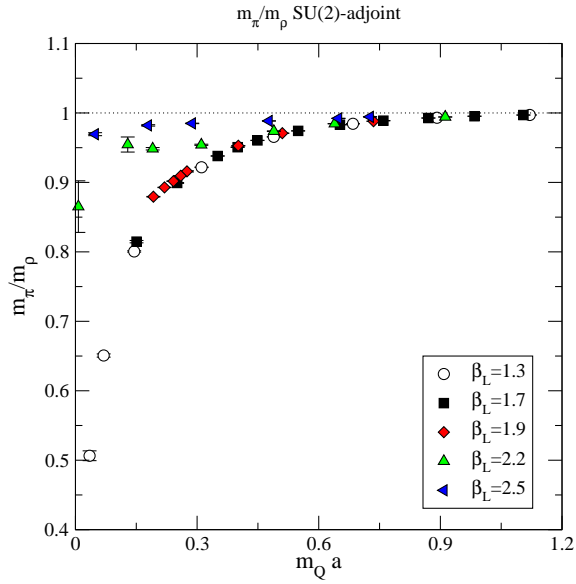}
\caption[a]{The ratio of the pseudoscalar and vector meson masses for the adjoint representation quarks at fixed values of $\beta_L$.  The near degeneracy of the masses at large $\beta_L$ is evident.}
\end{center}
\label{rho_pi_ratio_su2a}
\end{figure}

\begin{figure}
\begin{center}
\includegraphics[width=0.5\textwidth]{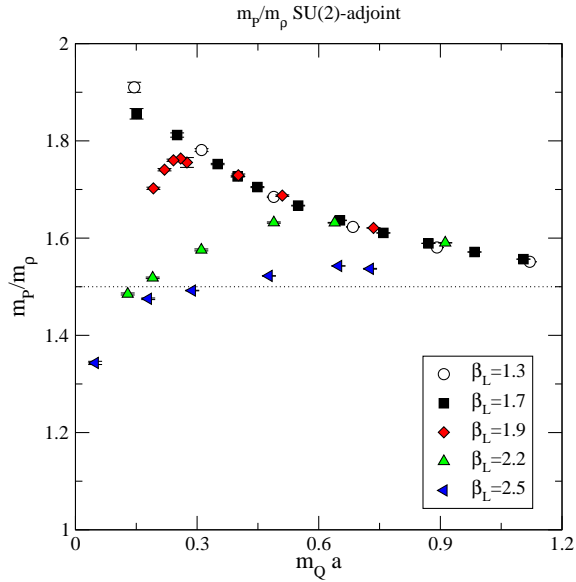}
\caption[a]{The ratio of the mass of the spin-1/2 three quark state (``proton'') and the vector meson mass, $m_P/m_\rho$.}
\end{center}
\label{proton_rho_ratio}
\end{figure}

As an alternative way to plot these results, in Fig.~\ref{rho_pi_ratio_su2a} we show the pseudoscalar-vector mass ratio $m_\pi/m_\rho$ for the theory with adjoint fermions.  The close degeneracy of the pseudoscalar and vector meson masses at large $\beta_L$ is evident.  In Fig.~\ref{proton_rho_ratio} we show the mass ratio of the three quark state with spin $1/2$ (``proton'') and the vector meson $\rho$.  We can observe that the mass ratio becomes flatter, and quite close to the value 1.5, as $\beta_L$ is increased.
The spin-3/2 three-quark state is almost degenerate with the spin-1/2 state.  

\begin{figure}
\begin{center}
\includegraphics[width=0.5\textwidth]{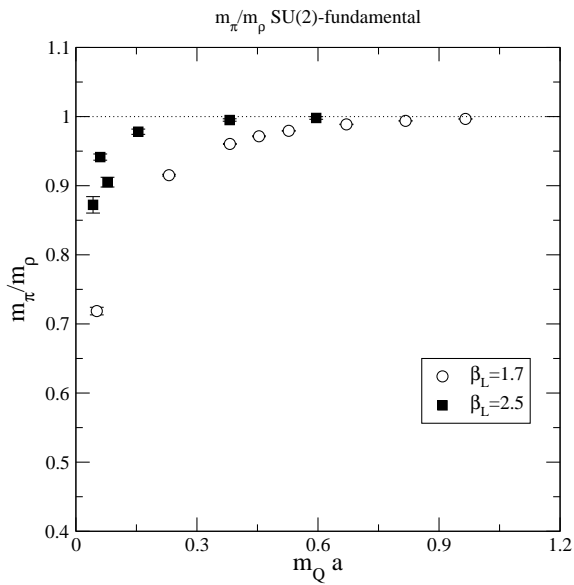}
\caption[a]{Same as Fig.~\ref{rho_pi_ratio_su2a} but for fundamental representation quarks.}
\end{center}
\label{rho_pi_ratio_su2}
\end{figure}

How to interpret these results?  At large $\beta_L$ (small bare coupling) all measured masses in the adjoint quark theory are approximately proportional to $m_Q$.  This behaviour is certainly compatible with a scale invariant behaviour at or near an infrared fixed point \cite{Sannino:2008pz}.  At small $\beta_L$ we observe chiral symmetry breaking, but we cannot quite reach the zero quark mass limit because of the first order phase transition at small $m_Q$.  If the theory has an exact IR fixed point, the chiral symmetry breaking is probably a lattice artifact not present in the continuum theory.  

However, this interpretation should be taken with care: qualitatively similar behaviour may arise even in the theory with fundamental fermions where the coupling is QCD-like, asymptotically free and running and the theory is in the chiral symmetry breaking phase. For comparison, in Fig.~\ref{rho_pi_ratio_su2} we show the ratio of the pseudoscalar and the vector meson masses in the SU(2) theory with fundamental fermions.  The result is qualitatively very similar to the adjoint representation case in Fig.~\ref{rho_pi_ratio_su2a}.  In this case the reason for this behaviour is clear: at small $\beta_L$ we observe the standard chiral symmetry breaking pattern, but at large $\beta_L$ the lattice volume becomes so small (lattice size $L < 1/\Lambda_{QCD}$) that the system becomes effectively deconfined, leading to behaviour which appears almost conformal at small $m_Q$. However, in this case we know that the chiral symmetry breaking appears at arbitrarily large $\beta_L$, provided that the lattice volume is large enough and the resolution of the measurements is sufficiently high.  Thus, the value of $\beta_L$ where the apparent chiral symmetry restoration happens should depend on the system size, with faster evolution of the running coupling giving stronger volume dependence.

In the simulations with the adjoint representation quarks we see no
systematic volume dependence in the location of the critical point where the chiral symmetry appears to become restored: in our range of volumes $8^4$ -- $32^4$, on $\beta_L=2.2$ lattices we observe no clear signs of chiral symmetry breaking (neglecting the $m_Q a \approx 0$ point in Fig.~\ref{rho_pi_ratio_su2a}), whereas at $\beta_L=1.9$ we already see the first order phase transition.  The stability of the
critical point shown in  Fig.~\ref{PD_su2a} supports the view that
it is a lattice artifact not present in the continuum theory.
If there is no chiral symmetry breaking on the weak coupling (large $\beta_L$) side of the critical point, the continuum theory does not possess it either.  This would be compatible with the infrared fixed point behaviour.  Naturally, our numerical results cannot exclude the possibility of chiral symmetry breaking even at large $\beta_L$; it may just evade detection within our accuracy and accessible volumes.  We discuss the implications of our results in more detail below.

\section{Conclusions and outlook}
\label{outlook}

We have investigated the strong coupling dynamics of SU(2) gauge theory with two fermion flavours in the adjoint representation of the gauge group. The aim of the study has been to determine whether this theory could provide a concrete example of (infrared) conformal or almost conformal theory. Phenomenologically such behaviour would be useful in the application to dynamical electroweak symmetry breaking, also known as walking technicolor, or some other beyond Standard model phenomenology like unparticle physics. We have determined the schematic phase diagram of the lattice theory and discussed its implications for continuum physics and also compared with the corresponding results in the case of fundamental representation fermions. Furthermore, we have analysed the spectrum of this theory. Even though the spectrum may be argued to present evidence in support of conformal behaviour, these arguments need to be treated with care. By comparing to similar measurements carried out for two flavours of fermions in the fundamental representation, we have shown that it is not possible to rule out the possibility of these behaviours being due to finite volume artifacts.  Definitive answer about the scale dependence of the coupling constant can only be provided by a dedicated measurement of the $\beta$-function of this theory.

Nevertheless, even without such measurement we can discuss how the results presented here can be contrasted with the evolution of the coupling. There are various options for asymptotically free coupling:

{\em I) QCD-like running coupling:~} the continuum limit is at $\beta_L \rightarrow\infty$, and there is spontaneous chiral symmetry breaking in the $m_Q\rightarrow 0$ limit.  However, at large $\beta_L$ the lattice volumes required to observe the chiral symmetry breaking are prohibitively large.  At too small volumes the theory may look almost conformal.  
However, because the chiral symmetry restoration is a finite volume effect, we can expect that at least some of the chiral observables should show clear volume dependence, which we do not observe here.

{\em II) IR fixed point:~} if $\beta_L$ is large enough, the theory
flows into the IR fixed point and there is no chiral symmetry breaking.  The formal continuum limit is still at $\beta_L\rightarrow\infty$.  The observed chiral symmetry breaking is a lattice artifact not connected to continuum physics (assuming $m_Q=0$).

{\em III) Walking coupling:~} while this belongs to the same universality class as case I (and thus has chiral symmetry breaking), 
depending on the degree of ``walking'' it interpolates between cases I and II.  Thus, it can be very hard to distinguish from either of the cases above. We can obtain this behaviour from case II by adding a small mass to the quarks, for example.

If we allow for $\beta(g)>0$, corresponding to the dashed-dotted line on the right side of the fixed point in Fig.~\ref{walk_beta}, there is also a further option corresponding to the free electric phase of the theory. In this case the theory is not asymptotically free and for small $\beta_L$, the theory flows into the IR fixed point and there is no chiral symmetry breaking. Since the continuum limit in this case is at $\beta_L \rightarrow 0$, our results show that the consistent continuum limit does not exist due to the observed onset of chiral symmetry breaking at $\beta_L\sim 2$. Of course, it is possible that chiral symmetry is broken in the free electric phase leading to even more exotic scenarios for the $\beta$-function than the three sketched in Fig.~\ref{walk_beta}: there could be multiple zeros, corresponding to nontrivial fixed points both in the infrared and ultraviolet. In such a case it would be possible to reach continuum also for $\beta>0$. Also, yet another logical possibility would be a single zero of second order, 
or even non-analytic behaviour of the $\beta$-function at the IR
fixed point.

To summarise, we have determined the lattice phase diagram and the excitation spectrum of SU(2) gauge field theory with two flavours of adjoint representation quarks.  This is a necessary step in the resolution of the strong coupling dynamics of the theory. While the results from the spectrum measurement alone are not sufficient to reliably resolve the type of the evolution of the coupling constant, the results are an important ingredient for future direct measurements of the full non-perturbative $\beta$-function. Particularly important phenomenon is chiral symmetry breaking which is well encoded into the spectrum of the theory. 

\acknowledgments
{We thank M.~Antola, K.~Kajantie and F.~Sannino for discussions and comments. JT and KR acknowledge the support of Academy of Finland grant 114371. AH acknowledges partial support by the NSF under grant number PHY-055375 and by US Department of Energy grant under contract DE-FG02-01ER41172. The simulations were performed at the Center for Scientific Computing (CSC), Finland, and at J\"ulich supercomputer center (JSC).
}


\begin{thebibliography}{199}

\bibitem{Sannino:2008ha}
  F.~Sannino, {\em{Dynamical Stabilization of the Fermi Scale: Phase Diagram of Strongly
 Coupled Theories for (Minimal) Walking Technicolor and Unparticles}},
  arXiv:0804.0182 [hep-ph].

\bibitem{Georgi:2007ek}
  H.~Georgi, {\em{Unparticle Physics}},
  Phys.\ Rev.\ Lett.\  {\bf 98}, 221601 (2007)
  [arXiv:hep-ph/0703260].

\bibitem{Georgi:2007si}
  H.~Georgi, {\em{Another Odd Thing About Unparticle Physics}},
  Phys.\ Lett.\  B {\bf 650}, 275 (2007)
  [arXiv:0704.2457 [hep-ph]].
  
\bibitem{Cheung:2007zza}
  K.~Cheung, W.~Y.~Keung and T.~C.~Yuan,
  {\em{Collider signals of unparticle physics}},
  Phys.\ Rev.\ Lett.\  {\bf 99}, 051803 (2007)
  [arXiv:0704.2588 [hep-ph]];
  F.~Sannino and R.~Zwicky,
  {\em{Unparticle \& Higgs as Composites}},
  arXiv:0810.2686 [hep-ph].

\bibitem{TC} 
S.~Weinberg,
{\em{Implications Of Dynamical Symmetry Breaking: An Addendum}},
Phys.\ Rev.\ D {\bf 19}, 1277 (1979);
L.~Susskind,
{\em{Dynamics Of Spontaneous Symmetry Breaking In The Weinberg-Salam Theory}},
Phys.\ Rev.\ D {\bf 20}, 2619 (1979).

\bibitem{Eichten:1979ah}
  E.~Eichten and K.~D.~Lane,
  {\em{Dynamical Breaking Of Weak Interaction Symmetries}},
  Phys.\ Lett.\  B {\bf 90}, 125 (1980).

\bibitem{Hill:2002ap}
C.~T.~Hill and E.~H.~Simmons,
{\em{Strong dynamics and electroweak symmetry breaking}},
Phys.\ Rept.\  {\bf 381}, 235 (2003) [Erratum-ibid.\  {\bf 390},
553 (2004)].

\bibitem{Holdom:1981rm}
B.~Holdom,
{\em{Raising The Sideways Scale}},
Phys.\ Rev.\ D {\bf 24}, 1441 (1981).

\bibitem{Yamawaki:1985zg}
K.~Yamawaki, M.~Bando and K.~i.~Matumoto,
{\em{Scale Invariant Technicolor Model And A Technidilaton}},
Phys.\ Rev.\ Lett.\  {\bf 56}, 1335 (1986).

\bibitem{Appelquist:an}
T.~W.~Appelquist, D.~Karabali and L.~C.~R.~Wijewardhana,
{\em{Chiral Hierarchies And The Flavor Changing Neutral Current Problem In
Technicolor}},
Phys.\ Rev.\ Lett.\  {\bf 57}, 957 (1986);
T.~Appelquist, A.~Ratnaweera, J.~Terning and
L.~C.~R.~Wijewardhana,
{\em{The phase structure of an SU(N) gauge theory with N(f) flavors}},
Phys.\ Rev.\ D {\bf 58}, 105017 (1998).

\bibitem{Ryttov:2007cx}
  T.~A.~Ryttov and F.~Sannino,
 {\em{Supersymmetry Inspired QCD Beta Function}},
  Phys.\ Rev.\  D {\bf 78}, 065001 (2008)
  [arXiv:0711.3745 [hep-th]].


\bibitem{Banks:1981nn}
  T.~Banks and A.~Zaks,
  {\em{On The Phase Structure Of Vector-Like Gauge Theories With Massless Fermions}},
  Nucl.\ Phys.\  B {\bf 196}, 189 (1982).


\bibitem{Jones:1977zr}
  D.~R.~T.~Jones,
  {\em{Charge Renormalization In A Supersymmetric Yang-Mills Theory}},
  Phys.\ Lett.\  B {\bf 72}, 199 (1977).

\bibitem{Brink:1982wv}
  L.~Brink, O.~Lindgren and B.~E.~W.~Nilsson,
  {\em{The Ultraviolet Finiteness Of The N=4 Yang-Mills Theory}},
  Phys.\ Lett.\  B {\bf 123}, 323 (1983).


\bibitem{Intriligator:1995au}
  K.~A.~Intriligator and N.~Seiberg,
  {\em{Lectures on supersymmetric gauge theories and electric-magnetic  duality}},
  Nucl.\ Phys.\ Proc.\ Suppl.\  {\bf 45BC}, 1 (1996)
  [arXiv:hep-th/9509066].

\bibitem{Sannino:2004qp}
  F.~Sannino and K.~Tuominen,
  {\em{Orientifold theory dynamics and symmetry breaking}},
  Phys.\ Rev.\  D {\bf 71}, 051901 (2005)
  [arXiv:hep-ph/0405209];
  D.~D.~Dietrich, F.~Sannino and K.~Tuominen,
  {\em{Light composite Higgs from higher representations versus electroweak
  precision measurements: Predictions for LHC}},
  Phys.\ Rev.\  D {\bf 72}, 055001 (2005)
  [arXiv:hep-ph/0505059];
  D.~D.~Dietrich and F.~Sannino,
  {\em{Conformal window of SU(N) gauge theories with fermions in higher dimensional representations}},
  Phys.\ Rev.\  D {\bf 75}, 085018 (2007)
  [arXiv:hep-ph/0611341].

\bibitem{Catterall:2007yx}
  S.~Catterall and F.~Sannino,
  {\em{Minimal walking on the lattice}},
  Phys.\ Rev.\  D {\bf 76}, 034504 (2007)
  [arXiv:0705.1664 [hep-lat]];
  L.~Del Debbio, A.~Patella and C.~Pica,
  {\em{Higher representations on the lattice: numerical simulations. SU(2) with
  adjoint fermions}},
  arXiv:0805.2058 [hep-lat];
  
\bibitem{Catterall:2008qk}
  S.~Catterall, J.~Giedt, F.~Sannino and J.~Schneible,
  {\em{Phase diagram of SU(2) with 2 flavors of dynamical adjoint quarks}},
  arXiv:0807.0792 [hep-lat].

\bibitem{Shamir:2008pb}
  Y.~Shamir, B.~Svetitsky and T.~DeGrand,
  {\em{Zero of the discrete beta function in SU(3) lattice gauge theory with color
  sextet fermions}},
  Phys.\ Rev.\  D {\bf 78}, 031502 (2008)
  [arXiv:0803.1707 [hep-lat]].


\bibitem{Appelquist:2007hu}
  T.~Appelquist, G.~T.~Fleming and E.~T.~Neil,
  {\em{Lattice Study of the Conformal Window in QCD-like Theories}},
  Phys.\ Rev.\ Lett.\  {\bf 100}, 171607 (2008)
  [arXiv:0712.0609 [hep-ph]].


\bibitem{Karsch:1998qj}
  F.~Karsch and M.~Lutgemeier,
  {\em{Deconfinement and chiral symmetry restoration in an SU(3) gauge theory
  with adjoint fermions}},
  Nucl.\ Phys.\  B {\bf 550}, 449 (1999)
  [arXiv:hep-lat/9812023].

\bibitem{Hietanen:2008vc}
  A.~Hietanen, J.~Rantaharju, K.~Rummukainen and K.~Tuominen,
  {\em{Spectrum of SU(2) gauge theory with two fermions in the adjoint
  representation}},
  PoS {\bf LATTICE2008}, 065 (2008)
  [arXiv:0810.3722 [hep-lat]].


\bibitem{Jansen:2008vs}
  K.~Jansen,
  {\em{Lattice QCD: a critical status report}},
  arXiv:0810.5634 [hep-lat].

\bibitem{Aoki:1986ua}
  S.~Aoki,
  {\em{Numerical Evidence For A Parity Violating Phase In Lattice QCD With Wilson
  Fermion}},
  Phys.\ Lett.\  B {\bf 190}, 140 (1987).

\bibitem{Sharpe:1998xm}
  S.~R.~Sharpe and R.~L.~Singleton,
  {\em{Spontaneous flavor and parity breaking with Wilson fermions}},
  Phys.\ Rev.\  D {\bf 58}, 074501 (1998)
  [arXiv:hep-lat/9804028];

\bibitem{DelDebbio:2008wb}
  L.~Del Debbio, M.~T.~Frandsen, H.~Panagopoulos and F.~Sannino,
  {\em{Higher representations on the lattice: perturbative studies}},
  JHEP {\bf 0806}, 007 (2008)
  [arXiv:0802.0891 [hep-lat]].

\bibitem{Sannino:2008pz}
  F.~Sannino,
  {\em{Conformal Chiral Dynamics}},
  arXiv:0811.0616 [hep-ph].

\end{thebibliography}
\end{document}